\documentclass{PoS}

\title{Hints of New Physics from Stars}

\ShortTitle{Hints of new physics from stars}

\author{\speaker{M. Giannotti}\\
	Physical Sciences, Barry University,
	11300 NE 2nd Ave., Miami Shores, FL 33161, USA\\
	E-mail: \email{mgiannotti@mail.barry.edu}}

\abstract{We review and update the current status of the observed anomalies in stellar cooling and their interpretation in terms of axions and axion like particles.}

\FullConference{38th International Conference on High Energy Physics\\
		3-10 August 2016\\
		Chicago, USA}

\begin{document}

\section{Introduction}

\begin{figure}[t]
	\begin{center}
		\includegraphics[width=0.45\linewidth]{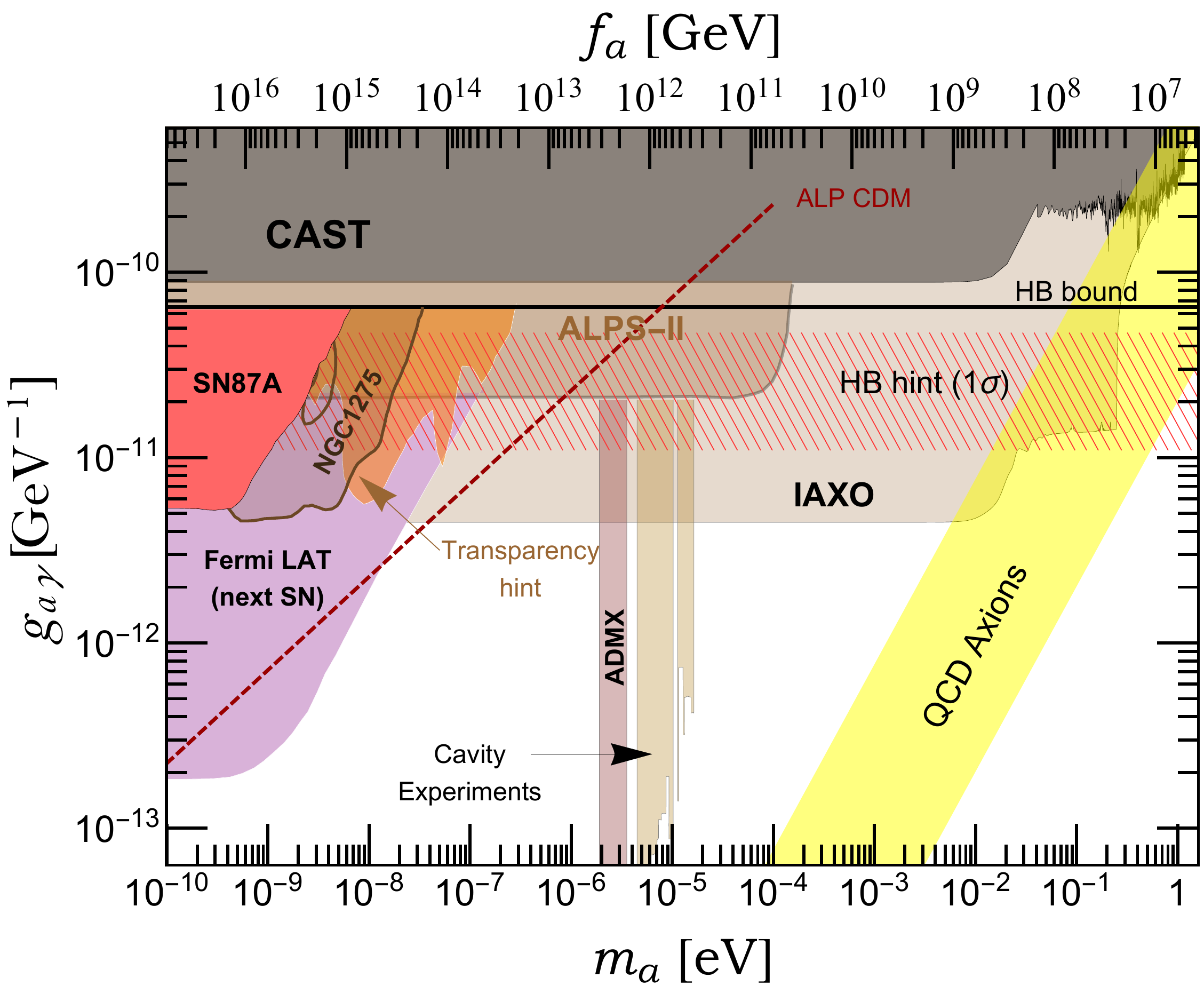}
		\hspace{0.3 cm}
		\includegraphics[width=0.43\linewidth, height=0.35\linewidth]{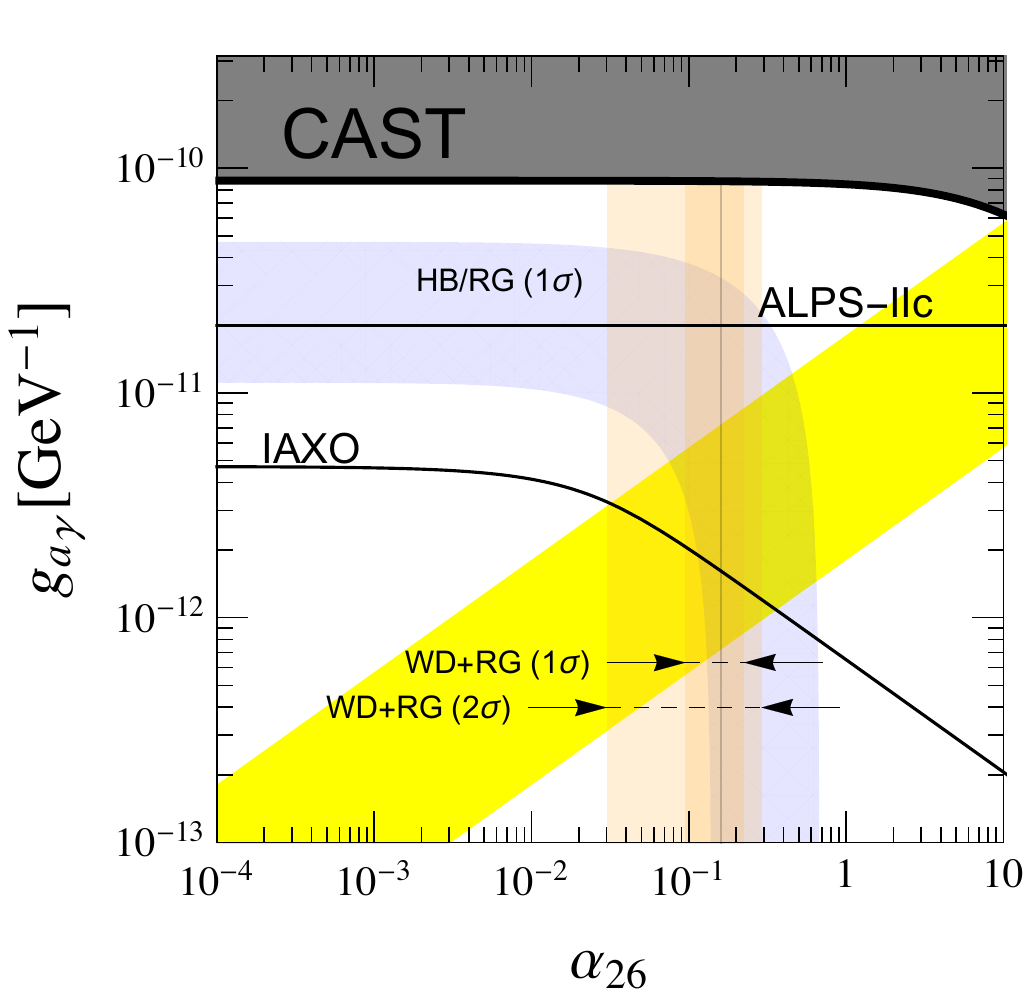}
		\caption{ALPs research status and hints. 
			The \textit{left panel} assumes ALPs interacting with photons only.
			The \textit{right panel} (from ref.~\cite{Giannotti:2015kwo}) shows the case of the ALPs interacting with photons and electrons. The electron coupling is expressed in terms of $ \alpha_{26} =(g_{ae}\times 10^{13})^2/4\pi$. 
			See text for more details.
			}
		\label{fig:sensitivity}
	\end{center}
\end{figure}
In spite of its enormous success, the standard model of particle physics does not provide a satisfactory explanation for several observations, leaving little doubt about the existence of \textit{new physics}.
Indeed, experimental searches for physics beyond the standard model are currently a major research area in physics and 
phenomenological hints to new physics have been playing a fundamental role in guiding the experimental efforts. 

Several new models (see, e.g., ref.~\cite{Ballesteros:2016euj,Ballesteros:2016xej,Berlin:2016bdv} for recent examples), which have attempted to address some of the fundamental problems unexplained by the standard model,
include axions or axion like particles (ALPs).
Axions~\cite{Weinberg:1977ma,Wilczek:1977pj} are light pseudoscalar particles predicted by the most widely accepted solution of the strong CP problem~\cite{Peccei:1977hh,Peccei:1977ur} and prominent dark matter candidates~\cite{Abbott:1982af,Dine:1982ah,Preskill:1982cy}.
Their interactions with photons and fermions are described by the Lagrangian terms
\begin{eqnarray}
L_{\rm int}=- \frac{1}{4}g_{a\gamma} \, a F_{\mu \nu} \tilde F^{\mu \nu} - \sum_{\rm fermions}g_{ai} \, a \overline \psi_i \gamma_5 \psi_i \,,
\label{eq:Lint}
\end{eqnarray}
where the coupling constants, $ g $, are inversely proportional to a phenomenological scale known as the Peccei-Quinn symmetry breaking scale. 
Moreover, in the so called QCD axion models, mass and interaction constants are directly proportional implying that low mass axions are also necessarily weakly interacting. 
The mass-coupling relation defines a band, known as the QCD axion band, indicated in yellow in Fig.~\ref{fig:sensitivity}.
It is important to remark, however, that the width of the band is somewhat arbitrary~\cite{DiLuzio:2016sbl} and that belonging to this band is not a requirement for the solution of the strong CP problem~\cite{Rubakov:1997vp,Berezhiani:2000gh,Gianfagna:2004je}.
More general models of pseudoscalar particles, known as ALPs, 
which couple to photons (and, possibly, to fermions) but do not satisfy the above mass-coupling relation, emerge naturally in various extensions of the Standard Model though, in general, their existence is not related to the strong CP problem~\cite{Ringwald:2012hr}. 

In recent years, more hints to new physics, mostly associated with axions or ALPs, have emerged from astrophysical observations. 
Among those, the seeming transparency of the universe to very high energy (E $ \gtrsim $ 100 GeV) gamma rays in the galactic and extragalactic medium~\cite{Horns:2012fx}\footnote{The lower mass section of the hinted region has been excluded by the non-observation of gamma rays from SN~1987A~\cite{Payez:2014xsa} and, more recently, by the search for spectral irregularities in the gamma ray spectrum of NGC~1275~\cite{TheFermi-LAT:2016zue}.} and some anomalous redshift-dependence of AGN gamma-ray spectra~\cite{Galanti:2015rda}. 
The recent discovery of peculiar periodic spectral modulations in some main sequence stars has also been attributed to axion-like particles~\cite{Tamburini:2016liq}.
\begin{table}[t]
	\begin{center}
		\begin{tabular}{ l  c  l  l l}
			\hline \hline
			\textbf{Observable} &	& \textbf{Stellar System}	&  \textbf{Proposed Solution(s)}  				& \textbf{References}	\\ \hline 
			Rate of	period		&	& WD Variables  			&  axion or ALP coupled to electrons;			& \cite{Corsico:2012ki,Corsico:2012sh,Corsico:2014mpa,Corsico:2016okh,Battich:2016htm}	\\ 
			change				& 	& 							&  neutrino magnetic moment; 					& \\ \hline 
			Shape of WDLF		&	& WDs					   	&  axion or ALP coupled to electrons; 			&\cite{Bertolami:2014noa,Bertolami:2014wua}\\ \hline 
			Luminosity of 		&	& Globular Clusters		   	&  axion or ALP coupled to electrons;			&\cite{Viaux:2013hca,Viaux:2013lha,Arceo-Diaz:2015pva}	\\ 
			the	RGB tip			&	&  (M5, $ \omega $-Centauri)							&  neutrino magnetic moment; 					&\\ \hline 
			R-parameter			&	& Globular Clusters  	  	&  axion or ALP coupled to photons;  			&\cite{Ayala:2014pea,OscarPatras}\\ \hline 
			B/R					&	& Open Clusters      		&  axion or ALP coupled to photons; 			&\cite{Skillman:2002aa,McQuinn:2011bb,Friedland:2012hj,Carosi:2013rla}\\ \hline 
			Neutron stars		&	& CAS A				 	  	&  axion or ALP coupled to neutrons.				&\cite{Leinson:2014ioa}\\ \hline \hline 
		\end{tabular}
		\caption{Summary of anomalous cooling observations. 
			All the anomalies have, individually, about 1-2~$ \sigma $ statistical significance, except for the last two for which the significance has not been quantified. }
		\label{tab:anomalies}   
	\end{center}
\end{table}

Additionally, several independent observations have shown anomalous behavior in the cooling of stellar objects~\cite{Ringwald:2015lqa,Giannotti:2015dwa,Giannotti:2015kwo}, as summarized in Tab.~\ref{tab:anomalies}. 
These include \textit{i)} several pulsating white dwarfs (WDs), in which the cooling efficiency was extracted from the rate of the period change; 
\textit{ii)} the WD luminosity function (WDLF), which describes the distribution of WD as a function of their brightness;
\textit{iii)} red giants branch (RGB) stars, in particular the luminosity of the tip of the branch; 
\textit{iv)} horizontal branch stars (HB) or, more precisely, the R-parameter, that is the ratio of the number of HB over RGB stars;
\textit{v)} helium burning supergiants, more specifically the ratio B/R of blue and red supergiants; 
and \textit{vi)} neutron stars.

What is remarkable in these observations is that, though each hint has \textit{per se} a small statistical significance, all the anomalous observations have shown an \textit{excessive cooling}
with respect to what expected from the standard stellar evolution theory, limiting the interpretation as random errors and indicating instead some systematic problem in our understanding of stellar cooling. 
All this led to the hypothesis that some novel light, weakly interacting particle(s) could be contributing to the energy loss in a way similar to how neutrinos do.
A recent comprehensive study of the various new physics options~\cite{Giannotti:2015kwo}, which included hidden photons, neutrino electromagnetic factors and minicharged particles, showed that axions and ALPs are the favorite candidates to explain the combined anomalies.
The 1~$ \sigma $ intervals on the axion/ALPs coupling with electrons and photons derived from this analysis are shown in Tab.~\ref{tab:WD_RGB_anomalies}.

The hint to a non-vanishing photon coupling was first observed in a study of the R-parameter in globular clusters~\cite{Ayala:2014pea,OscarPatras} . 
If the axion (or ALP) is coupled to photons only, its effect would be to accelerate the HB evolution while leaving essentially unchanged the RGB stage, and therefore to reduce the expected R-parameter. 
In fact, the Primakoff process, which is the most relevant production mechanism induced by the photon coupling in the stellar medium, is quite inefficient at the high density characterizing the RGB evolutionary stage.
Numerical analyses~\cite{Ayala:2014pea,OscarPatras} showed that a small axion-photon coupling would actually improve the agreement with the observed R-parameter, leading to the hint shown in Tab.~\ref{tab:WD_RGB_anomalies} and by the hashed region in the left panel of Fig.~\ref{fig:sensitivity}.
If the coupling to electrons cannot be neglected, the expected number of RGB stars could be modified by the Compton and Bremsstrahlung processes, particularly efficient in the RGB stage (which precedes the HB), and could therefore contribute to the change of the R-parameter~\cite{Giannotti:2015kwo}. 
In this case the hinted band in the $ g_{a\gamma}-\alpha_{26} $ parameter space is shown (blue region labeled HB/RG) in the right panel of Fig.~\ref{fig:sensitivity}.
Interestingly, a sufficiently large coupling to electrons could explain the R-parameter anomaly without any need for a coupling to photons, though this possibility is currently disfavored.

In the case of non-vanishing coupling with electrons, axions or ALPs  can also well explain the anomalies observed in WD and RGB stars, as shown in Tab.~\ref{tab:WD_RGB_anomalies}.
In particular, the Bremsstrahlung production rate, particularly efficient at the high densities characterizing the WD and RGB stars core, has a peculiar temperature dependence which makes the novel loss rate match the observed one with fairly good accuracy for all the different stellar systems. 
A combined analysis of these results indicates that the axion/ALP solution is favored at about 3$ \sigma $.
\begin{table}[t]
	\begin{center}
		\begin{tabular}{| l  l |  l l |}
			\hline 
			\textbf{observable} 					& \textbf{hint (1$ \sigma $)}		& \textbf{observable}			& \textbf{hint (1$ \sigma $)}					\\ \hline 
			WD G117 - B15A				&	 $ \alpha_{26}=1.89\pm 0.47 ~ $  	& 	WD R548 					&$ \alpha_{26}=1.84\pm 0.93 ~ $   	\\ 
			WD PG 1351+489 				&	 $ \alpha_{26}=0.36\pm 0.38~ $  	& 	WD L19-2   (113 s mode)		&	 $ \alpha_{26}=2.08\pm 1.35 ~ $   	\\ 
			WD L19-2  (192 s mode)		&	 $ \alpha_{26}=0.5\pm 1.2 ~ $   	& 	WDLF 						&$ \alpha_{26}=0.156 \pm 0.068~ $   		\\
			luminosity of RGB tip  		& $ \alpha_{26}=0.28^{+0.47}_{-0.24}~ $ &   R-parameter 				& $ g_{10}= 0.29\pm 0.18$	\\ \hline 
		\end{tabular}
		\caption{Hints at $ 1~ \sigma $ from stellar cooling with $ \alpha_{26} =(g_{ae}\times 10^{13})^2/4\pi$ and $ g_{10}=g_{a\gamma}\times 10^{10} $GeV. 
			In the case of the WD variables, the confidence intervals have been derived from a likelihood analysis of the data in~\cite{Corsico:2012sh,Corsico:2014mpa,Corsico:2016okh} and are not given in the original references.	
		The hint on the R-parameter shown here assumes no ALP-electron interaction.}
		\label{tab:WD_RGB_anomalies}   
	\end{center}
\end{table}

The cooling anomalies hint to a quite interesting region in the ALP parameter space. 
The HB hint, shown in the left panel of Fig.~\ref{fig:sensitivity}, and the combined hints from HB, WD and RGB, shown in the right panel, indicate a preferable region well accessible to the next generation axion helioscope experiment, IAXO~\cite{Irastorza:2011gs,Vogel:2013bta}, and also partially to the next generation axion shining through wall experiment ALPS II~\cite{Bahre:2013ywa}. 
At high mass these hinted regions overlap the QCD axion band for a range of parameters still partially accessible to IAXO while at lower masses they select ALPs which could provide the required CDM (area below the red dashed line in Fig.~\ref{fig:sensitivity}~\cite{Arias:2012az}).
Finally, at slightly lower masses the hinted region overlaps with the parameters required to explain the transparency hints~\cite{Horns:2012fx}.
Remarkably, the entire region hinted by both cooling anomalies and transparency would be easily accessible to the FERMI large area telescope in case of a future galactic SN~\cite{Meyer:2016wrm}. 

The possibility that the cooling anomalies may actually be hinting to new physics is an exciting, if speculative, prospect, with several running or planned experiments having a significant discovery potential in the favored regions. 
The negative results in the searches for a massive dark matter candidate could indicate that new physics may be hidden elsewhere, and stars could be hinting at its presence.

\end{document}